\documentclass[superscriptaddress,showpacs,preprint,prd]{revtex4}

\usepackage{latexsym}
\usepackage{graphicx}

\newcommand{\nc}{\newcommand}

\nc{\be}[1]{\begin{equation}\mbox{$\label{#1}$}}
\nc{\bea}[1]{\begin{eqnarray} \mbox{$\label{#1}$}}
\nc{\Section}[2]{\section{#2}\label{#1}}
\nc{\Bibitem}[1]{\bibitem{#1}}
\nc{\Label}[1]{\label{#1}}

\nc{\eea}{\end{eqnarray}}
\nc{\ee}{\end{equation}}

\nc{\bdm}{\begin{displaymath}}
\nc{\edm}{\end{displaymath}}
\nc{\dpsty}{\displaystyle}
\nc{\bc}{\begin{center}}
\nc{\ec}{\end{center}}
\nc{\ba}{\begin{array}}
\nc{\ea}{\end{array}}
\nc{\bab}{\begin{abstract}}
\nc{\eab}{\end{abstract}}
\nc{\btab}{\begin{tabular}}
\nc{\etab}{\end{tabular}}
\nc{\bit}{\begin{itemize}}
\nc{\eit}{\end{itemize}}
\nc{\ben}{\begin{enumerate}}
\nc{\een}{\end{enumerate}}
\nc{\bfig}{\begin{figure}}
\nc{\efig}{\end{figure}}

\nc{\arreq}{&\!=\!&}
\nc{\arrmi}{&\!-\!&}
\nc{\arrpl}{&\!+\!&}
\nc{\arrap}{&\!\!\!\approx\!\!\!&}
\nc{\non}{\nonumber}
\nc{\align}{\!\!\!\!\!\!\!\!&&}

\def\lsim{\; \raise0.3ex\hbox{$<$\kern-0.75em
      \raise-1.1ex\hbox{$\sim$}}\; }
\def\gsim{\; \raise0.3ex\hbox{$>$\kern-0.75em
      \raise-1.1ex\hbox{$\sim$}}\; }

\nc{\DOT}{\hspace{-0.08in}{\bf .}\hspace{0.1in}}
\nc{\Laada}{\hbox {$\sqcap$ \kern -1em $\sqcup$}}
\nc\loota{{\scriptstyle\sqcap\kern-0.55em\hbox{$\scriptstyle\sqcup$}}}
\nc\Loota{{\sqcap\kern-0.65em\hbox{$\sqcup$}}}
\nc\laada{\Loota}
\nc{\qed}{\hskip 3em \hbox{\BOX} \vskip 2ex}

\nc{\real}{{\rm I \! R}}
\nc{\Z}{{\sf Z \!\!\! Z}}
\nc{\complex}{{\rm C\!\!\! {\sf I}\,\,}}
\def\bigid{\leavevmode\hbox{\small1\kern-3.8pt\normalsize1}}
\def\id{\leavevmode\hbox{\small1\kern-3.3pt\normalsize1}}
\nc{\slask}{\!\!\!/}
\nc{\bis}{{\prime\prime}}
\nc{\pa}{\partial}
\nc{\na}{\nabla}
\nc{\ra}{\rangle}
\nc{\la}{\langle}
\nc{\goto}{\rightarrow}
\nc{\swap}{\leftrightarrow}

\nc{\EE}[1]{ \mbox{$\cdot10^{#1}$} }
\nc{\abs}[1]{\left|#1\right|}
\nc{\at}[2]{\left.#1\right|_{#2}}
\nc{\norm}[1]{\|#1\|}
\nc{\abscut}[2]{\Abs{#1}_{\scriptscriptstyle#2}}
\nc{\vek}[1]{{\rm\bf #1}}
\nc{\integral}[2]{\int\limits_{#1}^{#2}}
\nc{\inv}[1]{\frac{1}{#1}}
\nc{\dd}[2]{{{\partial #1}\over{\partial #2}}}
\nc{\ddd}[2]{{{{\partial}^2 #1}\over{\partial {#2}^2}}}
\nc{\dddd}[3]{{{{\partial}^2 #1}\over
    {\partial #2 \partial #3}}}
\nc{\dder}[2]{{{d #1}\over{d #2}}}
\nc{\ddder}[2]{{{d^2 #1}\over{d {#2}^2}}}
\nc{\dddder}[3]{{d^2 #1}\over
    {d #2 d #3}}
\nc{\dx}[1]{d\,^{#1}x}
\nc{\dy}[1]{d\,^{#1}y}
\nc{\dz}[1]{d\,^{#1}z}
\nc{\dl}[1]{\frac{d\,^{#1}l}{(2\pi)^{#1}}}
\nc{\dk}[1]{\frac{d\,^{#1}k}{(2\pi)^{#1}}}
\nc{\dq}[1]{\frac{d\,^{#1}q}{(2\pi)^{#1}}}

\nc{\bfT}{{\bf T }}

\nc{\cA}{{\cal A}}
\nc{\cB}{{\cal B}}
\nc{\cD}{{\cal D}}
\nc{\cE}{{\cal E}}
\nc{\cG}{{\cal G}}
\nc{\cH}{{\cal H}}
\nc{\cL}{{\cal L}}
\nc{\cO}{{\cal O}}
\nc{\cT}{{\cal T}}
\nc{\cN}{{\cal N}}
\nc{\cR}{{\cal R}}
\nc{\rvac}[1]{|{\cal O}#1\rangle}
\nc{\lvac}[1]{\langle{\cal O}#1|}
\nc{\rvacb}[1]{|{\cal O}_\beta #1\rangle}
\nc{\lvacb}[1]{\langle{\cal O}_\beta #1 |}
\nc{\bb}{\bar{\beta}}
\nc{\bt}{\tilde{\beta}}
\nc{\ctH}{\tilde{\cal H}}
\nc{\chH}{\hat{\cal H}}
\nc{\1}{\aa}
\nc{\2}{\"{a}}
\nc{\3}{\"{o}}
\nc{\4}{\AA}
\nc{\5}{\"{A}}
\nc{\6}{\"{O}}
\nc{\al}{\alpha}
\nc{\g}{\gamma}
\nc{\Del}{\Delta}
\nc{\eps}{\epsilon}
\nc{\lam}{\lambda}
\nc{\Om}{\Omega}
\nc{\ve}{\varepsilon}
\nc{\mn}{{\mu\nu}}
\nc{\vp}{\varphi}

\nc{\rf}[1]{(\ref{#1})}
\nc{\nn}{\nonumber \\*}
\nc{\bfB}{\bf{B}}
\nc{\bfv}{\bf{v}}
\nc{\bfx}{\bf{x}}
\nc{\bfy}{\bf{y}}
\nc{\vx}{\vec{x}}
\nc{\vy}{\vec{y}}
\nc{\oB}{\overline{B}}
\nc{\oI}{\overline{I}}
\nc{\oR}{\overline{R}}
\nc{\rar}{\rightarrow}
\nc{\ti}{\times}
\nc{\slsh}{\hskip-5pt/}
\nc{\sm}{Standard~Model~}
\nc{\MP}{M_{\rm Pl}}
\nc{\tp}{t_{\rm Pl}}

\nc{\pmin}{p_{\rm min}}
\nc{\pmax}{p_{\rm max}}
\nc{\fo}{f_0}
\nc{\foi}{f_{0,i}\,}
\nc{\fop}{f_0^P}
\nc{\fou}{f_0^U}

\nc{\eff}{{\rm eff}}
\nc{\MT}{M_{\rm T}}
\nc{\ML}{M_{\rm L}}
\nc{\kk}{\vek{k}}
\nc{\pp}{{\rm p}}
\nc{\half}{{1\over 2}}
\nc{\w}{\omega}
\nc{\uhat}{\hat{U}_\w}

\nc{\etal}{\mbox{\it et al.\,}}
\nc{\ie}{{\it i.e.\,}}
\nc{\eg}{{\it e.g.\,}}
\nc{\trh}{T_{\rm RH}}
\nc{\ad}{{a'\over a}}
\nc{\bd}{{b'\over b}}
\nc{\Rd}{{R'\over R}}
\nc{\diag}{{\textrm{diag}}}
\nc{\mato}[1]{\tilde{#1}}
\nc{\sech}{\textrm{sech}}
\nc{\I}{\textrm{I}}
\nc{\II}{\textrm{II}}
\nc{\III}{\textrm{III}}
\nc{\vev}[1]{\langle #1 \rangle}
\nc{\hyp}{\,\; F_{1{\hskip -16pt}2}{\hskip 11pt}}
\nc{\brhom}{\overline{\rho}_M}
\nc{\brho}{\overline{\rho}}
\nc{\rhob}{\overline{\rho}}
\nc{\Pb}{\overline{P}}
\nc{\bH}{\overline{H}}

\nc{\lcdm}{$\Lambda$CDM }

\def\my{\hbox{\large$\bigcirc$\hspace{-0.58cm}
\raise.5ex\hbox{\tiny{o o}}\kern-.62em
\lower.5ex\hbox{--}}\ }

\nc{\row}{\my\my\my\my\my\my\my}
 
\nc{\mytilde}[1]{{\hskip 2.2pt}$\tilde{}${\hskip -2.2pt}#1}

\begin{document}

\title{The Integrated Sachs-Wolfe effect as a probe of non-standard cosmological evolution}

\author{Tuomas Multam\"{a}ki}
\email{tuomas@nordita.dk}
\affiliation{NORDITA, Blegdamsvej 17, DK-2100, Copenhagen, DENMARK}
\author{\O ystein Elgar\o y}
\email{oelgaroy@astro.uio.no}
\affiliation{Institute of theoretical astrophysics, University of Oslo, 
P.O. Box 1029, 0315 Oslo, NORWAY}

\date{December ??, 2003}

\begin{abstract}
The Integrated Sachs-Wolfe effect is studied in non-standard cosmologies. By considering
flat universes with a non-fluctuating dark energy component, it is shown how the quadrupole
power can be suppressed by atypical evolution of the scale factor. For example, a brief
period of non-standard evolution at a high redshift can suppress the quadrupole significantly.
The effect on the overall normalization of the CMB power spectrum is also discussed.
Non-standard cosmologies can affect the overall normalization significantly and enhance
the primordial fluctuations. The possibility of constraining
such non-standard models with CMB and independent measures of $\sigma_8$, is
considered.
\end{abstract}

\pacs{98.80.Es}

\preprint{NORDITA-2003-99 AP}

\maketitle

\section{Introduction}

There is mounting evidence that we are living in a universe dominated 
by a dark energy component, acting as a source of gravitational repulsion 
causing late-time acceleration of the expansion rate.  Early hints 
came from the classical test of using the magnitude-redshift relationship 
with galaxies as standard candles \cite{solheim}, but the reality 
of cosmic acceleration was not taken seriously until the 
magnitude-redshift relationship was measured recently using high-redshift 
supernovae type Ia (SNIa) \cite{riess,perlmutter}.  Cosmic acceleration 
requires a contribution to the energy density with negative pressure, 
the simplest possibility being a cosmological constant.  Independent 
evidence for a non-standard contribution to the energy budget of 
the universe comes from e.g. the combination of CMB and large-scale 
structure:  the position of the first peak in the CMB is consistent 
with the universe having zero spatial curvature, which means that 
the energy density is equal to the critical density.  However, 
large-scale structure shows that the contribution of standard 
sources of energy density, whether luminous or dark, is only a 
fraction of the critical density.  Thus, an extra, unknown component 
is needed to explain the observations \cite{efstathiou}.

The primary anisotropies in the cosmic microwave background (CMB) 
radiation have their origin in effects in the recombination era 
when photons and baryons decoupled.  In particular, the by now 
familiar pattern of peaks in the CMB power spectrum is interpreted 
as acoustic oscillations in the photon-baryon plasma prior to last 
scattering.  In addition to the primary anisotropies, secondary 
anisotropies may arise as the photons travel from last scattering 
at a redshift $z\sim 1100$ to us.  One such source of secondary 
anisotropies is CMB photons climbing in and out of evolving 
gravitational potential wells \cite{sw,rees}, this is the so-called 
Integrated Sachs-Wolfe (ISW) effect.   
In an Einstein-de Sitter 
universe, the gravitational potential is time-independent and hence 
there is no ISW effect.  In contrast, in the $\Lambda{\rm CDM}$ model 
the gravitational potential will start to decay once the cosmological 
constant starts dominating the expansion.  This will produce an 
extra contribution to the CMB anisotropies on large angular scales 
\cite{crittenden1}.  Additionally, large-scale anisotropies are caused 
by gravitational potential wells present at the last-scattering surface, 
the ordinary Sachs-Wolfe effect \cite{sw}. 

Working in the conformal Newtonian gauge, the perturbed metric 
can be written as 
\begin{equation}
ds ^2 = a^2 (\eta)[d\eta^2(1+2\Psi)-(1+2\Phi)d{\bf x}^2], 
\label{eq:metric}
\end{equation}
where $\eta$ is the conformal time, $d\eta = dt/a$, 
and $\Psi = -\Phi$ in the absence of anisotropic stress, and can 
be interpreted as the Newtonian gravitational potential.  Photons 
travelling to us from the last scattering surface obey the collisionless 
Boltzmann equation 
\begin{equation}
\frac{\partial}{\partial \eta}(\Theta + \Phi) + n^i \frac{\partial }
{\partial x^i}(\Theta + \Psi) = 0, 
\label{eq:boltzmann}
\end{equation}
where $\Theta(\eta,{\bf x},{\bf n})$ is the fractional temperature 
perturbation observed in the direction ${\bf n}$ on the sky at 
the conformal time $\eta$ and position ${\bf x}$.  After a Fourier 
and a Legendre transform,
\begin{eqnarray}
\Theta(\eta,{\bf x},{\bf n}) &=&\int \frac{d^3 k}{(2\pi)^3} 
\Theta(\eta,{\bf k},{\bf n})e^{i{\bf k}\cdot{\bf x}} 
\label{eq:fourier} \\ 
\Theta(\eta,{\bf k},{\bf n})&=&\sum_{\ell = 0}^\infty 
(-i)^\ell (2\ell + 1)\Theta_\ell(\eta,k)P_\ell(\mu), \label{eq:legendre}
\end{eqnarray}
where $\mu = \hat{{\bf k}}\cdot{\bf n}$, and $P_\ell(x)$ is the 
Legendre polynomial of degree $\ell$,  one can show that the 
solution to the Boltzmann equation on large scales is given by 
\begin{eqnarray}
\Theta_\ell(\eta,k)&=& [\Theta_0(\eta_{\rm r},k)+\Psi(\eta_{\rm r},k)]
j_\ell(k(\eta_0-\eta_{\rm r})) \nonumber \\ 
&+& \int_{\eta_{\rm r}}^{\eta_0}d\eta e^{-\tau(\eta)}(\Psi'
-\Phi')j_\ell(k(\eta_0-\eta)), \label{eq:solution}
\end{eqnarray}
where $\eta_{\rm r}$ is the conformal time at recombination, $\eta_0$ is 
the conformal time at the present epoch, $\tau$ is the optical 
depth, and the $j_\ell$ are the spherical Bessel functions.  
Primes denote derivatives with respect to conformal time. 
The CMB power spectrum is given by 
\begin{eqnarray}
C_\ell &=& 4\pi \int \frac{d^3 k}{(2\pi)^3}\langle |\Theta_\ell(\eta_0,k)|^2 
\rangle \nonumber \\
&=& 4\pi \int\frac{d^3 k}{(2\pi)^3} |
[\Theta_0(\eta_{\rm r},k)+\Psi(\eta_{\rm r},k)]
j_\ell(k(\eta_0-\eta_{\rm r})) \nonumber \\ 
&+& \int_{\eta_{\rm r}}^{\eta_0}d\eta e^{-\tau(\eta)}(\Psi'
-\Phi')j_\ell(k(\eta_0-\eta))|^2. \label{eq:cmbpower}
\end{eqnarray}
The first term corresponds to the Sachs-Wolfe effect, the second term 
is the contribution from the Integrated Sachs-Wolfe effect.  

Since the gravitational potential is related to the matter density  
through the Poisson equation, one should expect correlations between the 
ISW effect and the local matter distribution \cite{crittenden1}.  
Several detections of this effect has been reported in the last year, 
based on correlating the WMAP measurements of the CMB anisotropies 
\cite{wmap1,wmap2,wmap3,wmap4,wmap5,wmap6} 
with various tracers of the mass distribution 
\cite{boughn,nolta,fosalba1,fosalba2,scranton,afshordi}.    
In \cite{boughn}, the CMB was cross correlated with the 
hard X-ray background observed by the HEAO-1 satellite \cite{heao1} 
and the NVSS survey of radio galaxies \cite{nvss}, and a 2-3$\sigma$ 
detection of an ISW signal was reported.  Nolta {\it et al.} \cite{nolta} 
also investigated the cross-correlation with the NVSS catalogue within 
the $\Lambda{\rm CDM}$ model, and found 2$\sigma$ evidence for a 
non-zero cosmological constant.  The WMAP data was cross-correlated 
with the APM Galaxy survey \cite{apm} in \cite{fosalba1}, 
and a 98.8\% detection at the largest angular scales was found.  
Furthermore, in \cite{fosalba2,scranton} a detection of the ISW 
signal was reported from the cross-correlation of WMAP with 
various samples of galaxies from the Sloan Digital Sky Survey 
\cite{kevork}.   Finally, the cross-correlation with the 2MASS 
galaxy survey was investigated in \cite{afshordi}, resulting in a  
$2.5\sigma$ detection of the ISW signal, consistent with the 
expected value for the concordance $\Lambda{\rm CDM}$ model.  
These results give important evidence for the presence of a dark 
energy component in the universe, independent of the SNIa 
results, but are not at the level of precision where they can accurately 
pin down the properties of the dark energy.  

A feature of the WMAP results \cite{wmap1,wmap2,wmap3,wmap4,wmap5,wmap6} 
which has attracted a lot of attention is the lack of power on large scales.  
In \cite{wmap5} the two-point correlation of the WMAP data shows an 
almost complete lack of signal on angular scales greater than 60 degrees, 
and according to \cite{wmap5} the probability of finding such a result 
in the overall best fitting $\Lambda{\rm CDM}$ model is about 
$1.5\times 10^{-3}$.  In the power spectrum, this lack of large-scale 
power is evident in the low value of the quadrupole and, to a lesser 
extent, of the octopole.  This has spurred a great deal of interest, inspiring 
several authors to introduce exotic physics to explain this feature 
\cite{george2,contaldi,cline,feng,dedeo}.   In this paper we investigate 
whether a novel dark energy component can explain the apparent puzzle 
in the CMB data through its influence on the Integrated Sachs-Wolfe effect.  
Note, however, that the statistical significance of the apparent lack 
of large-scale power in the WMAP data has been called into question, 
and the actual discrepancy with the concordance $\Lambda{\rm CDM}$ model 
may only be at the level of a few percent \cite{tuomas,costa,george3}. 

\section{ISW effect in a flat universe}

When no anisotropic stress is present, $\Psi = - \Phi$.  For adiabatic, 
scale-invariant  fluctuations, the Sachs-Wolfe contribution is given by 
$\Theta_0(\eta_{\rm r},k)+\Psi(\eta_{\rm r},k)= \Psi(\eta_{\rm r},k)/3$ 
with $\Psi^2(\eta_{\rm r},k)\propto k ^{-3}$.  Furthermore, writing 
$\Psi(\eta,k)=5\Psi(\eta_{\rm r},k)f(\eta)/3$, one finds that 
at large scales, the power spectrum in a flat universe with a dark energy component 
that does not fluctuate is described by \cite{kofman}
\be{isw}
C_l={A^2\over 100 \pi l (l+1)}K_l^2,
\ee
where
\bea{Kl2}
K_l^2 & = &  200 l (l+1) \int_0^\infty {dk\over k}\Big[{1\over 10}j_l(k\eta_0)+\int_0^{\eta_0} d\eta\ {df\over d\eta} j_l(k(\eta_0-\eta))\Big]^2\nonumber\\
& \equiv & 200 l(l+1)\widetilde{K}_l^2.
\eea
Here $A$ is the overall normalization of the primordial fluctuations, $j_l$ is the 
spherical Bessel function and $\eta=\int_0^{t}dt/a(t)$ is the conformal time.  
In equation (\ref{Kl2}) we have approximated $\eta_{\rm r}=0$, which for large
multipoles is a reasonable approximation \cite{kofman}.
We have also ignored any effects arising from a finite optical depth.
Looking at Eq. (\ref{Kl2}) and recalling that $\int_0^\infty dk j_l(k\eta_0)^2/k=1/2l(l+1)$,
one sees that, if $df/d\eta$ vanishes, $K_l$ is a constant.

Note that the above equations hold only for a dark energy component that does not
collapse gravitationally. The question of perturbations of the dark energy is very central 
when calculating the magnitude of the ISW effect. In dark energy models with $w>-1$, 
the inclusion of perturbations
increase large scale power whereas for $w<-1$ the opposite occurs \cite{weller}.
The effect for the $w<-1$ models can be dramatic, where for $w=-2$ the inclusion 
of dark energy perturbations reduces the power by more than a factor of two \cite{weller}.
One should hence be cautious when calculating the ISW effect a particular model 
to make sure that possible perturbations are correctly accounted for.

The function $f$ is defined as
\be{fdef}
f(t)\equiv 1-{\dot{a}\over a^2}\int^t_0 dt'\ a(t').
\ee
In calculating the ISW effect, we need the derivative with respect to conformal time:
\be{dfdn}
{df\over d\eta}={1\over a^2}\Big(3({a'\over a})^2-{a''\over a}\Big)\int_0^{\eta} a^2\ d\eta-{a'\over a},
\ee
where now $a=a(\eta)$.

From Eq. (\ref{dfdn}) it is straightforwardly seen that if $f=f_0=const.$, then 
\be{f0const}
a \sim t^{{1\over f_0}-1}.
\ee
In an EdS universe, $a\sim t^{2/3}$, and hence $f=3/5$ as can be seen directly from Eq. (\ref{fdef})
or from Eq. (\ref{f0const}). Therefore there is no ISW effect in a universe that expands as a 
power-law, \eg a universe dominated by matter or radiation. A flat \lcdm universe, with 
\be{lcdm}
a(t)=a_0 \Big({\Omega_M\over\Omega_{\Lambda}}\Big)^{\frac 13} \sinh^{\frac 23}(\frac 32 \sqrt{\Omega_{\Lambda}}H_0t),
\ee
on the other hand, does not expand as a power law and hence $f$ varies with time leading
to a non-zero ISW effect.

By expanding the square in Eq. (\ref{Kl2}),  we note that  $\widetilde{K}_l^2$ is made of three terms:
two positive terms and a term whose overall sign is indeterminate and depends on the
evolution of the universe. It is precisely this term that allows one, at least in principle,
to have less power at large multipoles compared to a \lcdm universe
(the possible significance of the cross term was also mentioned in \cite{contaldi}).
Of the three different $k$-integrals the first one can be readily evaluated,
\be{int1}
\int_0^\infty {dk\over k} j_l(k\eta_0)^2={1\over 2l(l+1)},
\ee
whereas the other two are of the form
\be{int2}
I_l(\eta_1,\eta_2)\equiv \int_0^\infty {dk\over k} j_l(k(\eta_0-\eta_1)) j_l(k(\eta_0-\eta_2))
\ee
and cannot evaluated as easily. For numerical calculations, it is useful to express
Eq. (\ref{int2}) in terms of hypergeometric functions:
\be{int3}
I_l(\xi)=2^{2l-1}\xi^l{l!(l-1)!\over (2l+1)!} \hyp (-\frac 12,l,\frac 32+l,\xi^2),
\ee 
where $\xi\equiv(\eta_0-\eta_1)/(\eta_0-\eta_2)<1$ (since the integral is symmetric
with respect to $\eta_1\rar\eta_2,\  \eta_2\rar\eta_1$, we can always choose $\xi$ to
be less than one).
This allows us to express $\widetilde{K}_l^2$ in terms of $I_l$ as
\bea{kl3}
\widetilde{K}_l^2 & = & {1\over 200}{1\over l(l+1)}+{1\over 5}\int_0^{\eta_0}d\eta\ {df\over d\eta}
I_l(1-\frac \eta\eta_0)+\int_0^{\eta_0}d\eta_1\ {df\over d\eta_1}\int_0^{\eta_0}d\eta_2\ {df\over d\eta_2} I_l(\xi).
\eea
In this form the integrals are numerically simple to compute as the numerical approximations
to hypergeometric functions are readily available.  We have checked that we can successfully 
reproduce the shape of the large scale multipoles (up to the accuracy of our analytical approximation), 
in the case where the dark energy component does not fluctuate compared to results from a full CMB code 
\cite{weller}. 

\begin{figure}[ht]
\begin{center}
\includegraphics[width=60mm,angle=-90]{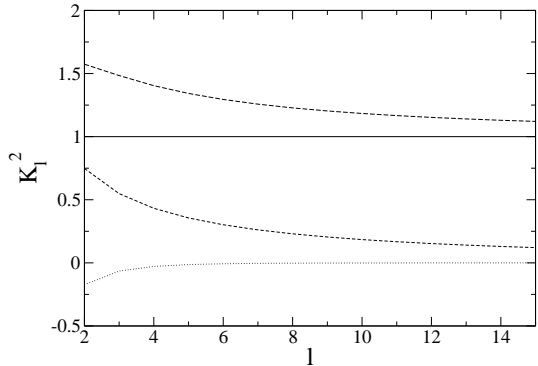}
\caption{$K_l^2$ in the \lcdm model: the solid curve is the SW-contribution, dotted line is the cross term, dashed line is the square term and the long-dashed line is the sum}\label{ipic}
\end{center}
\end{figure}

Calculating the value of $K_l^2$ at different multipoles is straightforwardly done
using Eq. (\ref{kl3}). As an example, we have plotted the different contributions
to the low multipoles in a \lcdm model with $\Omega_M=0.25$ in Fig.  \ref{ipic}.
From the figure we see that the cross term is negligible. In the \lcdm model it is hence
well justified to use the approximation 
$|\Delta^{SW}+\Delta^{ISW}|^2\approx (\Delta^{SW})^2+(\Delta^{ISW})^2$
when calculating the ISW effect.
Generally this does not have to be true, however, and one can wonder whether the ISW can be 
responsible for the observed low quadrupole. Also, one can pose the inverse problem: given a set 
of multipoles, what is $a(t)$ if there are only matter perturbations?

\section{Suppressing the large scale power with non-standard evolution}
To illustrate how one can reduce the large scale CMB power with the ISW effect
we consider a number of different models.

\subsection{Jump in $f(t)$}
As a simple toy model of the universe,  we consider an evolution history such that $f$ undergoes a jump 
at some high redshift:
\be{deltaf}
{df\over d\eta}=f_0\delta(\eta-\eta_c).
\ee
Physically this would require that the expansion law changes 
somewhere from the surface of last scattering until now, \eg there is a period after recombination
where the expansion goes as $a\sim t^\gamma,\ \gamma\neq 2/3$, and then resumes the
normal matter dominated behaviour.

Before and after the jump, $f$ is constant and the derivative vanishes.
Inserting {\it Ansatz} (\ref{deltaf}) into Eq. (\ref{kl3}) we find that 
\be{deltakl}
\widetilde{K}_l^2={1\over 2 l(l+1)}\Big({1\over 100}+f_0^2\Big)+
\frac 15 f_0 I_l(1-\eta_c/\eta_0).
\ee
From this we see that one can reduce power at a particular multipole to zero if
the condition
\be{zerocond}
l(l+1)I_l(1-\eta_c/\eta_0)>\frac 12
\ee
holds. However,  $I_l(\xi)<1/2l(l+1),\ \xi<1$ so that 
this is only possible if $\eta_c=0$ which means that $f_0=-1/10$.
But this choice leads to vanishing power at all of the multipoles considered as is obvious
from Eq. (\ref{Kl2}). 

Similarly, it is straightforward to see that given a jump at $\eta_c$, one
can reduce power at a given multipole, $l$, to
\be{klmin}
^{min}\widetilde{K}_l^2={1\over 100 l (l+1)}-{1\over 50}l(l+1)I_l(1-\eta_c/\eta_0).
\ee
However, we are interested in suppressing power at a given multipole {\it relative} to other
multipoles. In particular we wish to suppress $C_2$ with respect to a normalization scale
which we choose to be $C_{10}$.
Minimizing $K^2_2/K_{10}^2$ we find that $f_0=-1/10$ 
(this result does not depend on which two multipoles we are considering).
The choice $f_0=-0.1$ corresponds to a period where $f=3/5-0.1=0.5$, which from Eq. (\ref{f0const})
implies that during that time $a\sim t$. 

\begin{figure}[ht]
\begin{center}
\includegraphics[width=60mm,angle=-90]{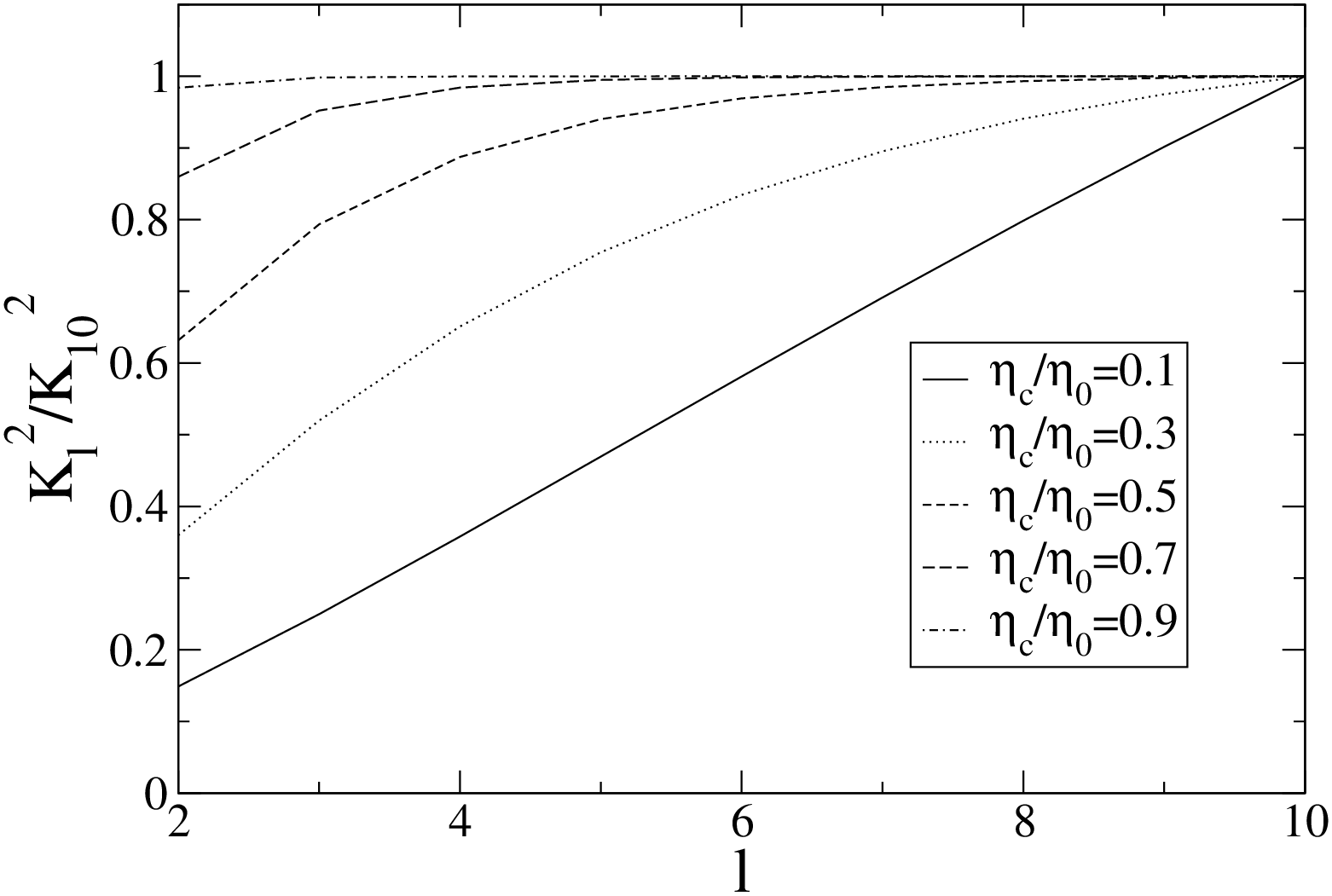}
\includegraphics[width=60mm,angle=-90]{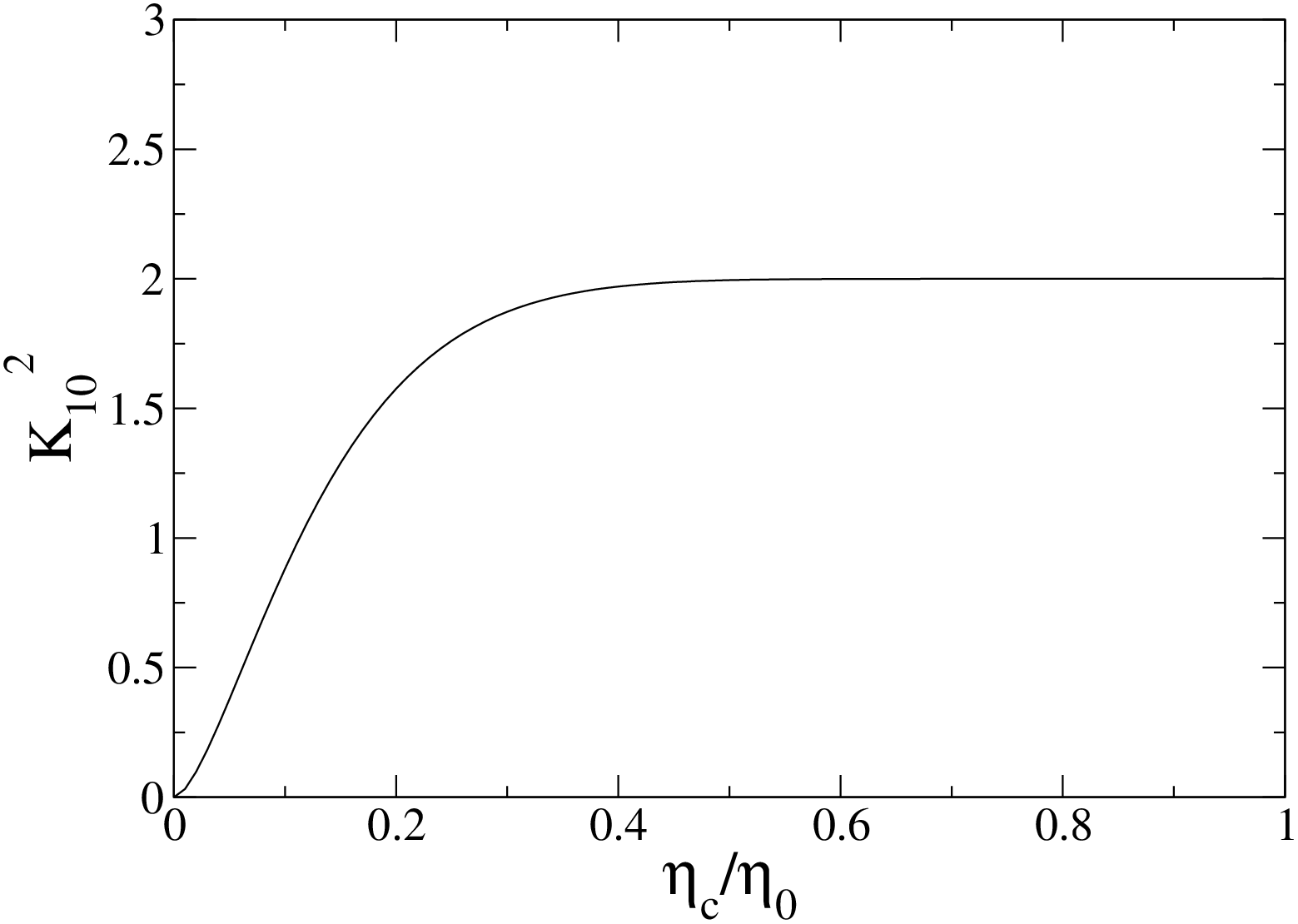}
\caption{$K_l^2$(left) and $K_{10}^2$ (right) for different values of $\eta_c/\eta_0$, $f_0=-0.1$.}
\label{deltapic}
\end{center}
\end{figure}

The effect of varying $\eta_c$ is shown in Fig. \ref{deltapic}. We see that the earlier
the transition happens, the stronger the suppression.
In the same figure we have plotted the normalization factor.

\subsection{Jump in $f(t)$ in a \lcdm universe}
On might be inclined to believe that a jump in $f$ at a very high redshift is
decoupled from the effects due to acceleration at low redshifts in a standard 
\lcdm cosmology. However, they are not decoupled but a cross term is present again and hence
one cannot extend the conclusions of the previous section to the \lcdm model.

Let us therefore consider a modification of the \lcdm cosmology by adding a jump to $f$ at some
redshift:
\be{ljump}
f(\eta)=f_{\Lambda}(\eta)+f_0\theta(\eta-\eta_c),
\ee
where $f_{\Lambda}$ is calculated in a \lcdm universe with $\Omega_M=0.25$.
Inserting this into Eq. (\ref{kl3}), we get
\be{ljumpkl}
\widetilde{K}_l^2=\widetilde{K}_l^{2,\Lambda}+{f_0^2\over 2l(l+1)}+\frac 15 f_0 I
l\Big(1-\frac\eta\eta_c\Big)+2f_0\int_0^{\eta_0}d\eta\ {df_{\Lambda}\over d\eta}I_l
\Big({\eta_0-\eta\over\eta_0-\eta_c}\Big),
\ee
where $\widetilde{K}_l^{2,\Lambda}$ is the \lcdm contribution.

\begin{figure}[ht]
\begin{center}
\includegraphics[width=60mm,angle=-90]{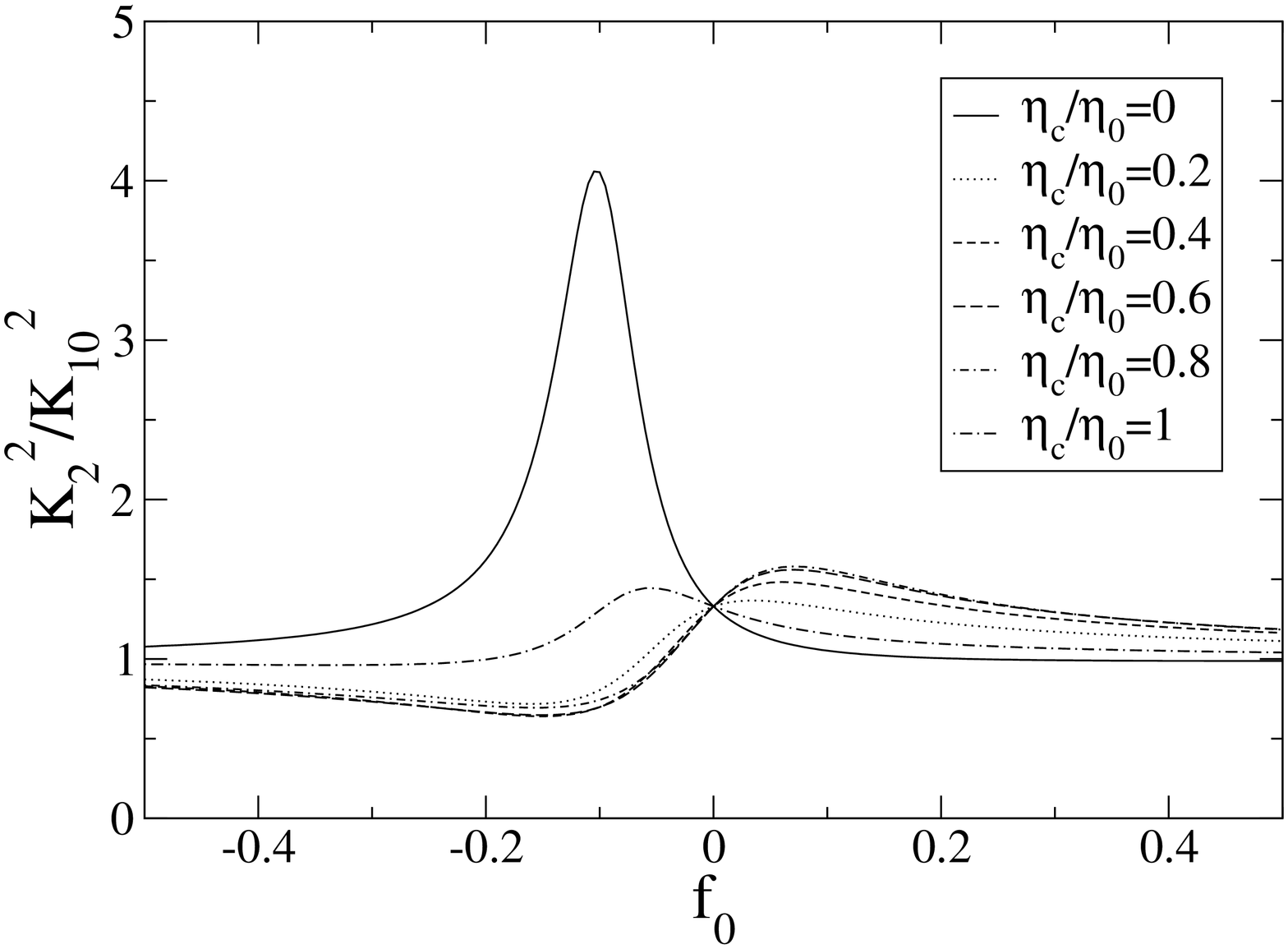}
\includegraphics[width=60mm,angle=-90]{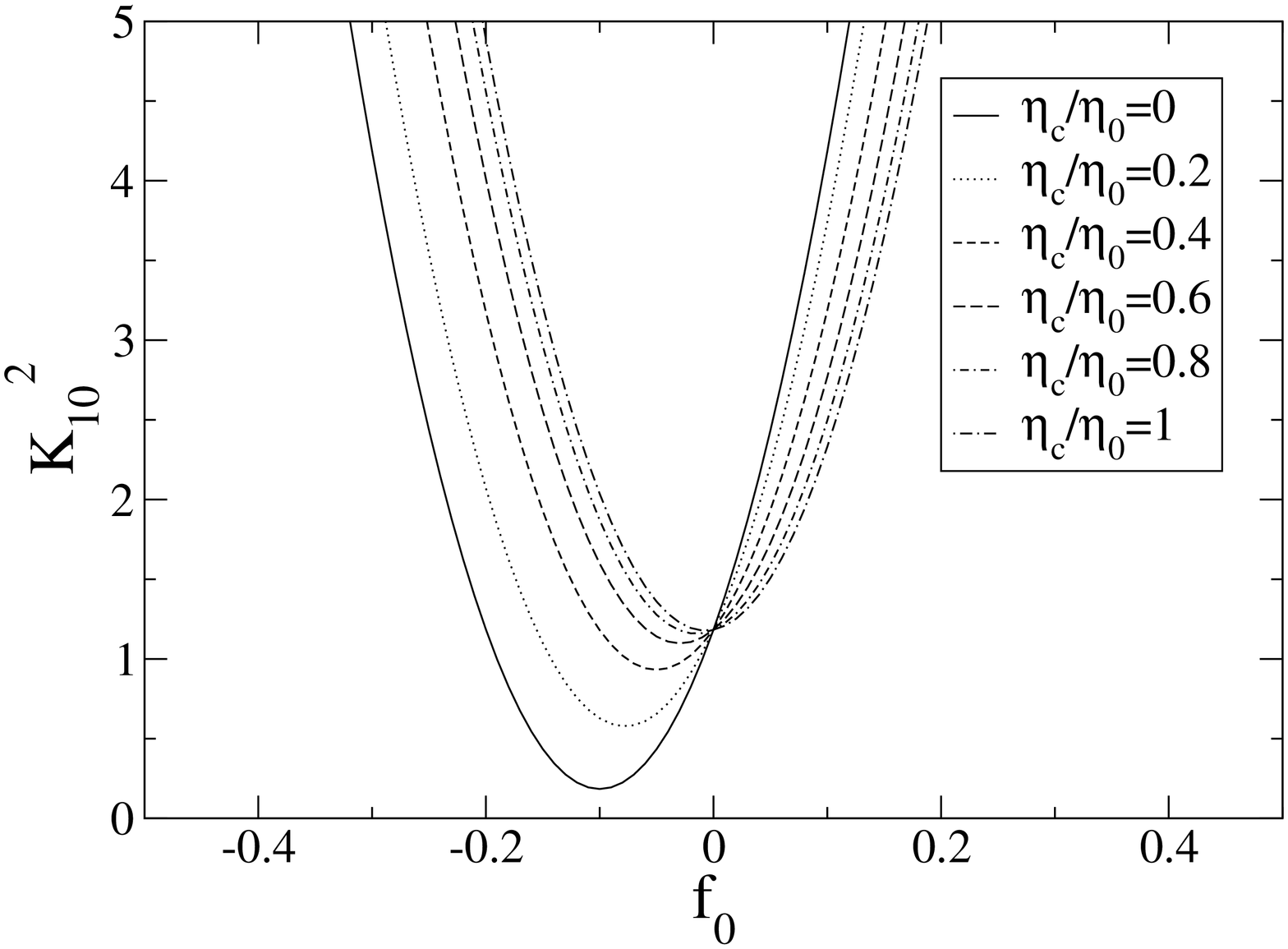}
\caption{The relative quadrupole power for different  $\eta_c$ (left) and the normalization
factor right) as a function on $f_0$}\label{lcdmpic}
\end{center}
\end{figure}

We have plotted the quadrupole power relative to the normalization scale in Fig. \ref{lcdmpic} for 
different values of $f_0$ and $\eta_c$. From the figure we see that overall it is difficult to suppress
the relative quadrupole power by a large amount.
The maximum suppression can be found numerically,
\be{lcdmmin}
{K_2^2\over K_{10}^2}\approx 0.64,\ f_0\approx -0.15,\ {\eta_c\over\eta_0}\approx 0.67.
\ee
Hence, one cannot suppress the quadrupole power by a very significant amount by such 
a mechanism. Furthermore, the length of non-standard evolution required is long and the 
corresponding expansion, $a\sim t^8$, is significantly different from standard matter dominated behaviour.

\subsection{\lcdm with a crunch}

As a more realistic possibility, we consider a modified \lcdm model where the universe undergoes a 
period of non-standard growth sometime after recombination.
This can be straightforwardly modelled by
\be{thetalcdm}
{df\over d\eta}={df_{\Lambda}\over d\eta}+f_0\theta(\frac 12{\Delta\eta}-|\eta-\eta_c|),
\ee
where $\Delta\eta$ is the duration of the non-standard phase.
Inserting the{\it Ansatz} (\ref{thetalcdm}) into Eq. (\ref{kl3}), it is easy to see that
\bea{ljumpkl2}
\widetilde{K}_l^2 & = & \widetilde{K}_l^{2,\Lambda}+f_0^2\Big({1\over 2l(l+1)}-2I_l({\eta_0-\eta_1\over
\eta_0-\eta_2})\Big)+\frac 15 f_0 \Big(I_l(1-{\eta_1\over\eta_c})-I_l(1-{\eta_2\over\eta_c})\Big)\nonumber\\
& & + 2f_0\int_0^{\eta_0}d\eta\ {df_{\Lambda}\over d\eta}
\Big(I_l({\eta_0-\eta_1\over\eta_0-\eta})-I_l({\eta_0-\eta_2\over\eta_0-\eta})\Big),
\eea
where $\eta_1\equiv\eta_c-\Delta\eta/2,\ \eta_2\equiv\eta_c+\Delta\eta/2$.

\begin{figure}[ht]
\begin{center}
\includegraphics[width=60mm,angle=-90]{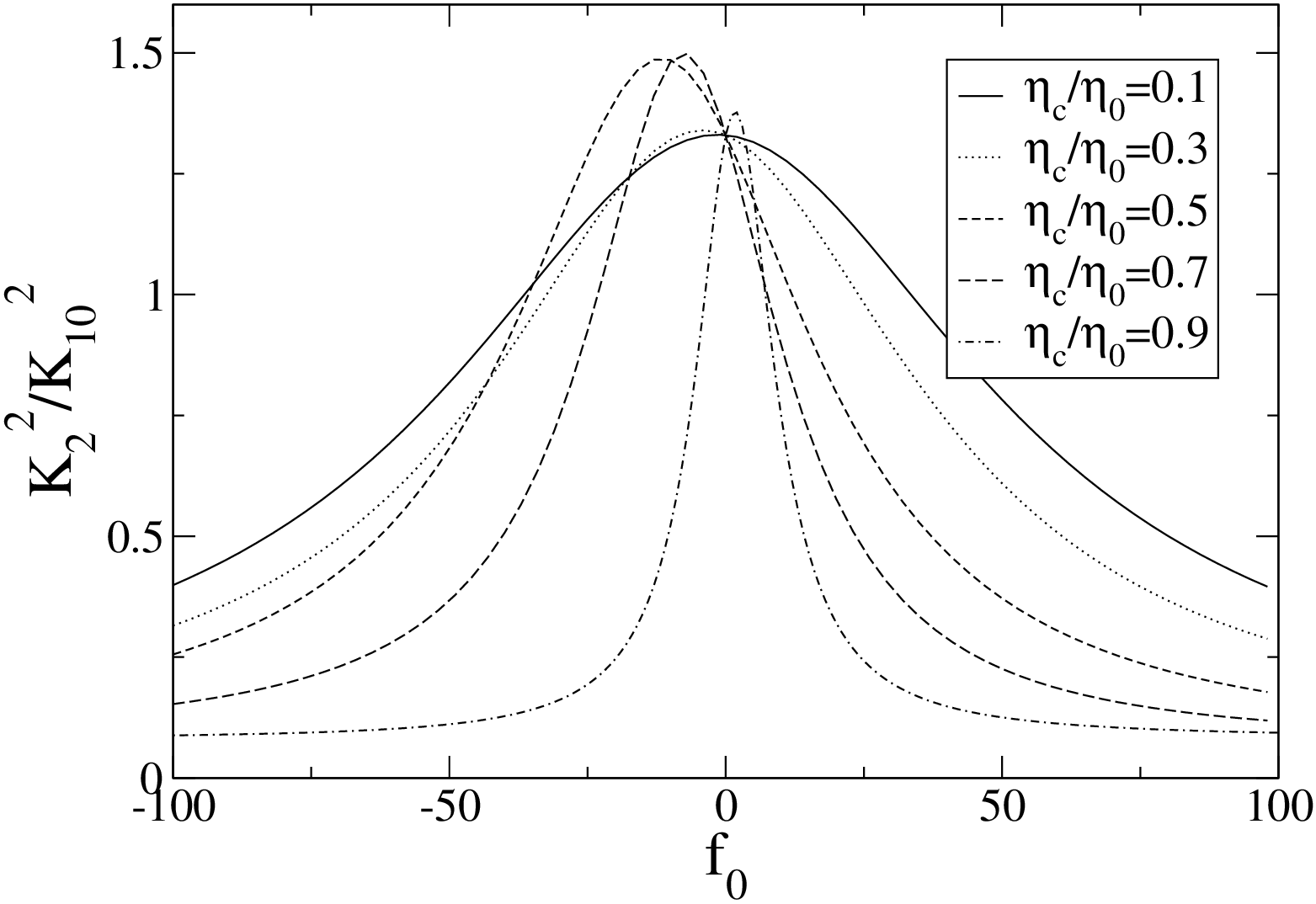}
\includegraphics[width=60mm,angle=-90]{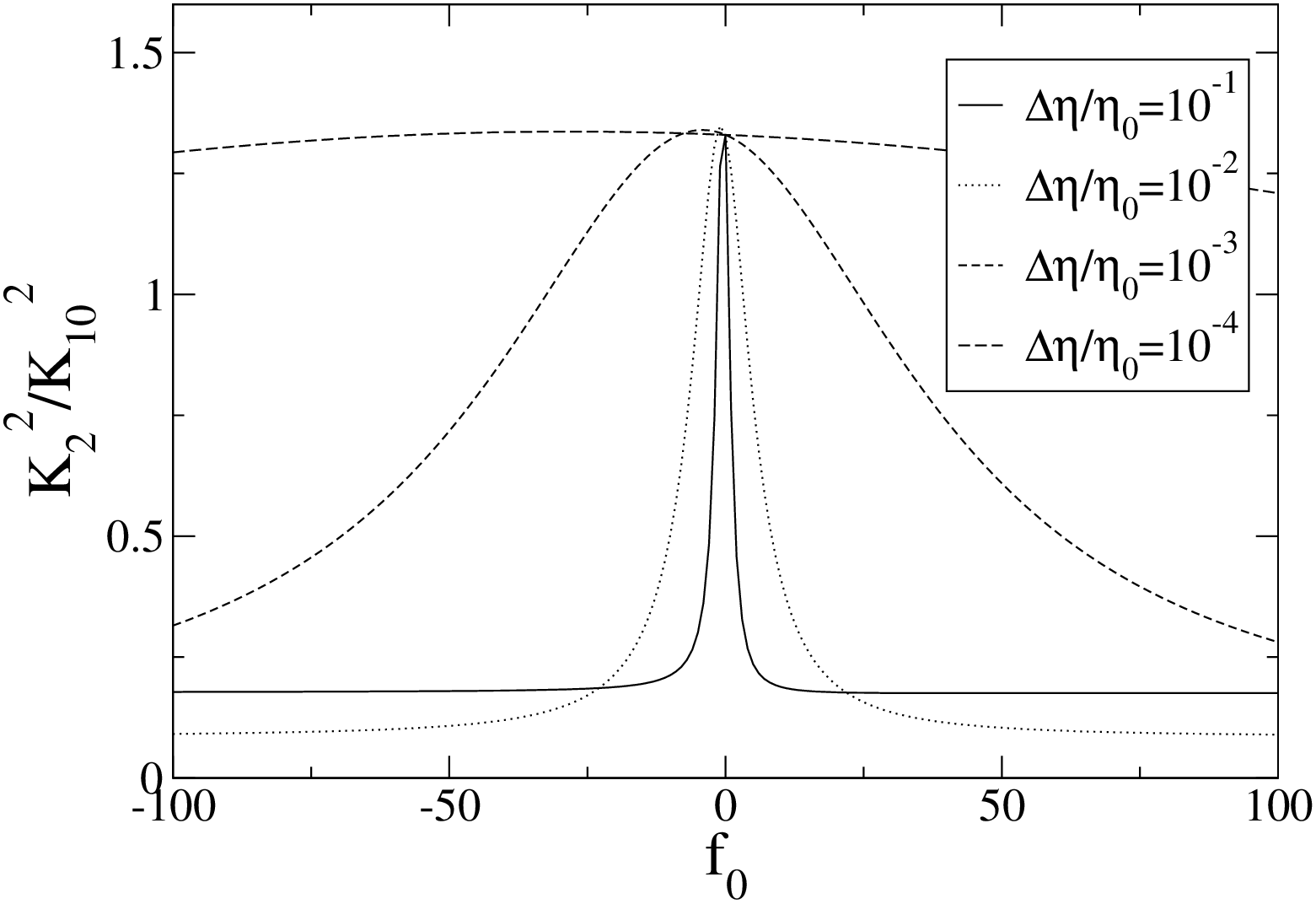}
\caption{The relative quadrupole power as a function of $f_0$ for different values of $\eta_c/\eta_0$
($\Delta\eta/\eta_0=10^{-3}$) (left) and $\Delta\eta/\eta_0$ ($\eta_c/\eta_0=0.3$) (right)}\label{ddlcdm}
\includegraphics[width=60mm,angle=-90]{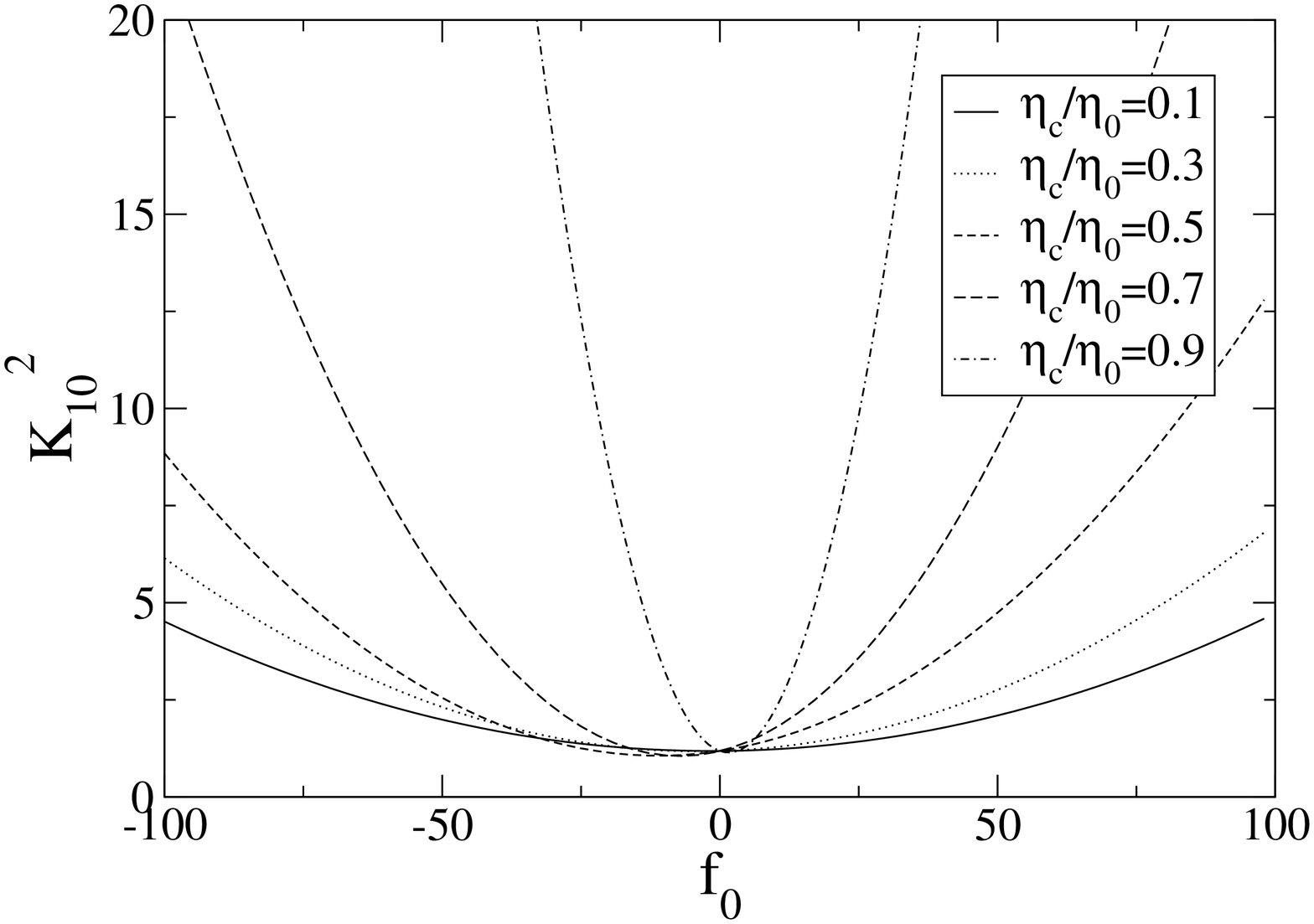}
\includegraphics[width=60mm,angle=-90]{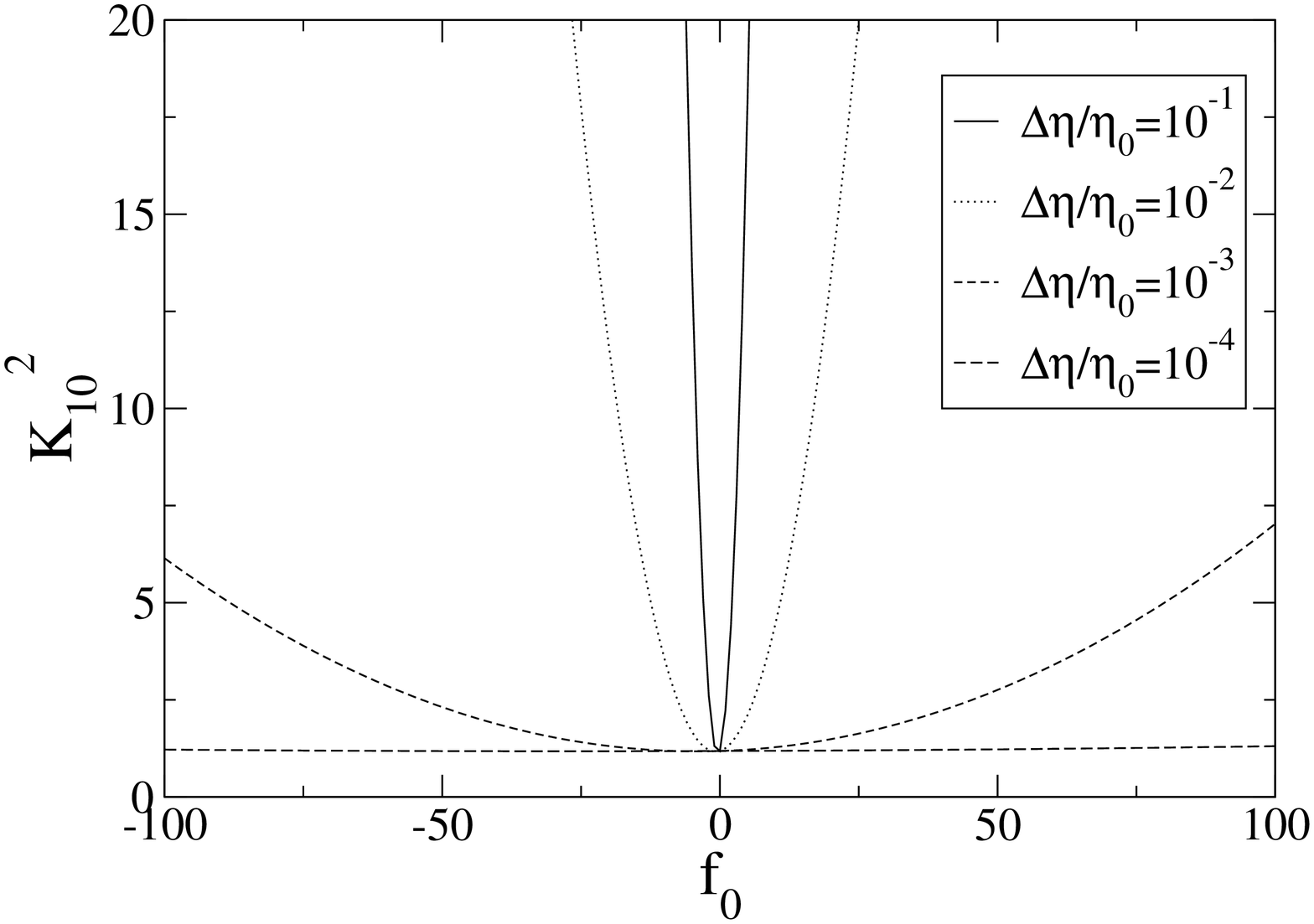}
\caption{Normalization as a function of $f_0$ for different values of $\eta_c/\eta_0$ ($\Delta\eta/\eta_0=10^{-3}$) 
(left) and $\Delta\eta/\eta_0$ ($\eta_c/\eta_0=0.3$) (right)}
\label{ddlcdm2}
\end{center}
\end{figure}

Now we have three free parameters in the model, $f_0$ and the time and duration of the
period of non-standard growth, $\eta_c$ and $\Delta\eta$. We have explored the parameter space
numerically and the result are shown in Figs \ref{ddlcdm} and \ref{ddlcdm2}.
From Fig. \ref{ddlcdm} we see that the relative quadrupole power as a function of $f_0$
has a peaked shape. The position of the peak is determined by $\eta_c$ whereas
$\Delta\eta$ determines the width. Note that these figures indicate that one can reduce
the relative quadrupole power quite significantly. Looking at the overall normalization, shown in  
Fig. \ref{ddlcdm2}, we see that as $|f_0|$ is increased the overall power increases rapidly.
As a general feature we see that in order to decrease the relative quadrupole power,
we need to increase $f_0$ which in turn increases $K_{10}^2$ and hence changes the overall
CMB normalization.

Given $K_{10}^2$, one can consider how much the quadrupole can be suppressed in this model.
Scanning the parameters we have found that in order to reduce the quadrupole power by half, 
we typically need to change the overall normalization by a factor of $3-4$. 

In order to explain the lack of large scale power while not changing the history of the universe
too radically, we would like to have a very short period of non-standard
growth occurring at a high enough redshift. Choose say, $\Delta\eta/\eta_0=0.001$,
and approximate the evolution of the universe by \lcdm for purposes of calculating redshifts. 
The value of $f_0$ is set by requiring that $K_{10}^2$ is not bigger than $\sim 2$.
\begin{figure}[ht]
\begin{center}
\includegraphics[width=60mm,angle=-90]{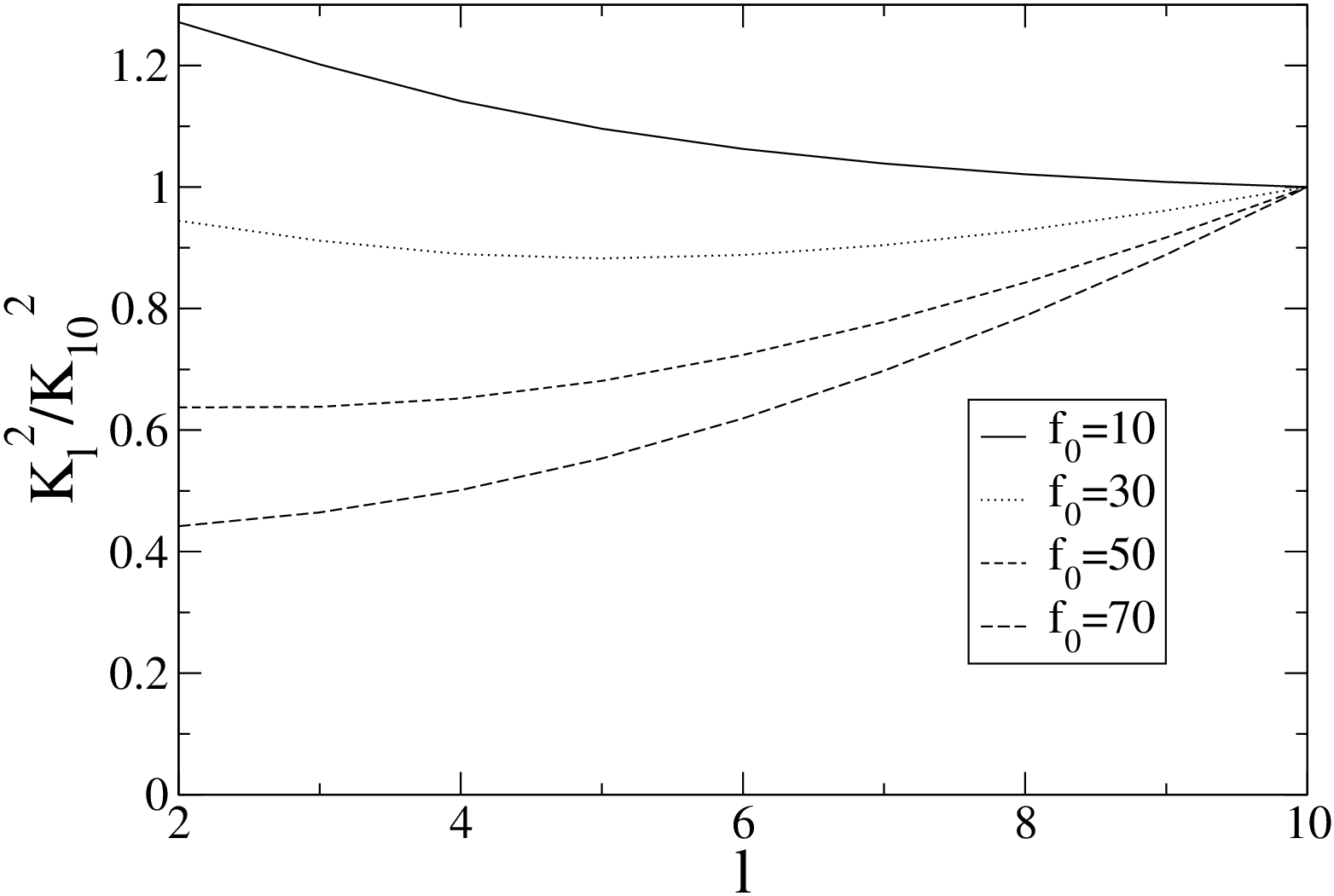}
\includegraphics[width=60mm,angle=-90]{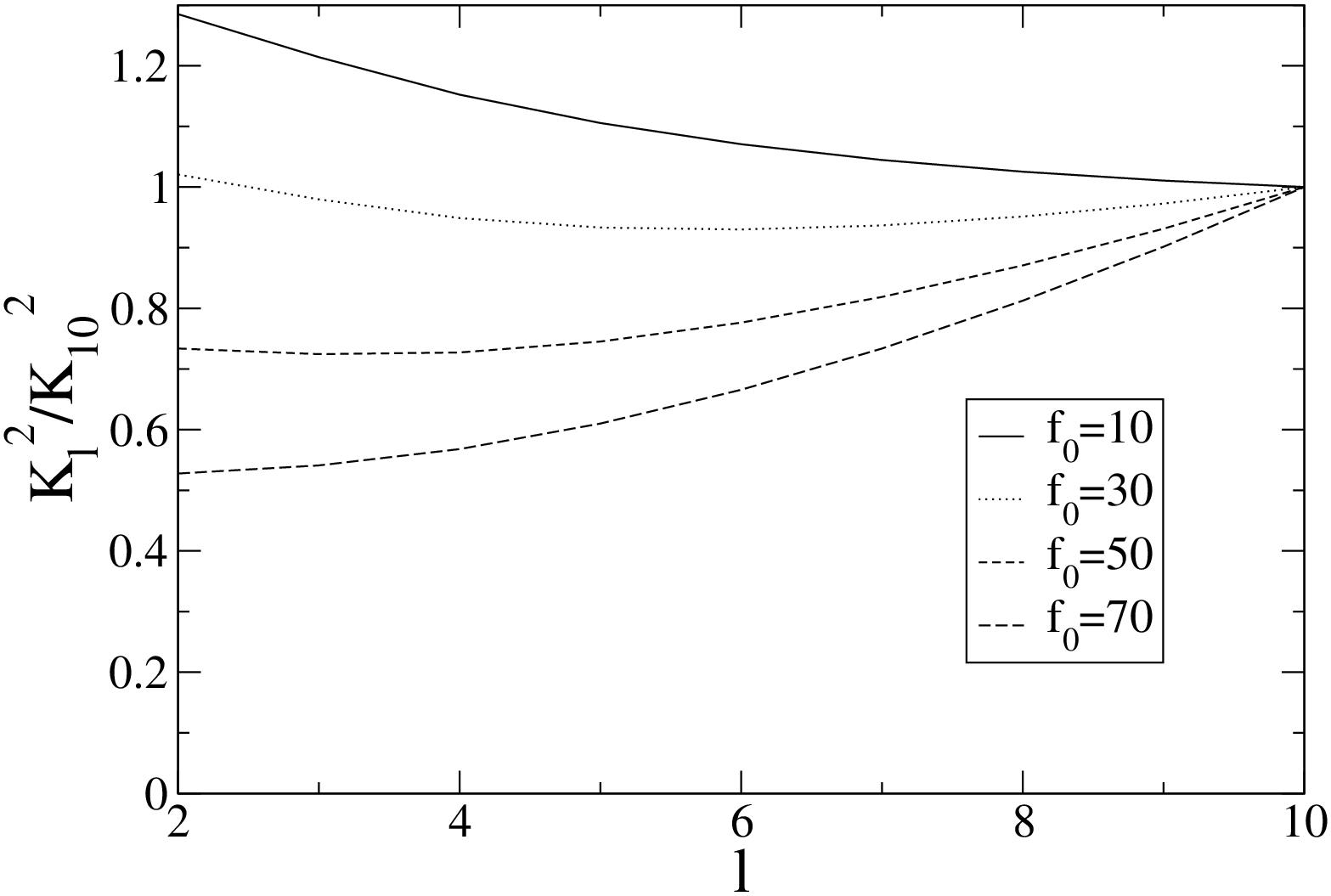}
\caption{Relative power at large scales with $z_c\sim 10$ (left) and $z_c\sim 30$ (right)}
\label{ddlcdm4}
\end{center}
\end{figure}  
The multipole power for such a model is shown in Fig. \ref{ddlcdm4} for $z_c\sim 10,\ 30$.
We see that one can suppress the quadrupole effectively by a short period of non-standard
growth. It is also evident that the later this occurs, the larger the effect.
Physically, the parameter values correspond to a short period during which the scale factor 
evolves as $a\sim 1/t$, as can be seen from Eq. (\ref{f0const}). Hence, the universe shrinks briefly!

\subsection{Modified Friedmann equations}
As a final example on the possibility of suppressing the relative quadrupole power, we consider
models where the Friedmann equation is modified from the ordinary one. Such models can be interesting
from the point of view of the ISW effect since linear fluctuations grow differently from in
the standard scenario and one can possibly use the ISW effect to constrain such models
\cite{copeland}. 
As a particular example, consider the Modified Polytropic Cardassian (MPC) model \cite{mpc}, 
in which the Friedmann equation is modified in such a way that SNIa observations are fit by having universe
filled only with matter,
\be{card}
H^2={8\pi G\over 3}\rho_M\Big(1+({\rho_M\over\rho_c})^{q(n-1)}\Big)^{1/q},
\ee
where $\rho_M$ is the energy density of matter, $\rho_c$ is the critical density at which 
the non-standard terms begin to dominate and $q>0,\ n<2/3$  are parameters.
The growth of linear perturbations can be radically different from the \lcdm model in such models \cite{mgm}.
With large $q$ there is a period where linear growth is enhanced compared to the \lcdm model and
hence the MPC-models are potentially interesting for suppressing the quadrupole.

\begin{figure}[ht]
\begin{center}
\includegraphics[width=60mm,angle=-90]{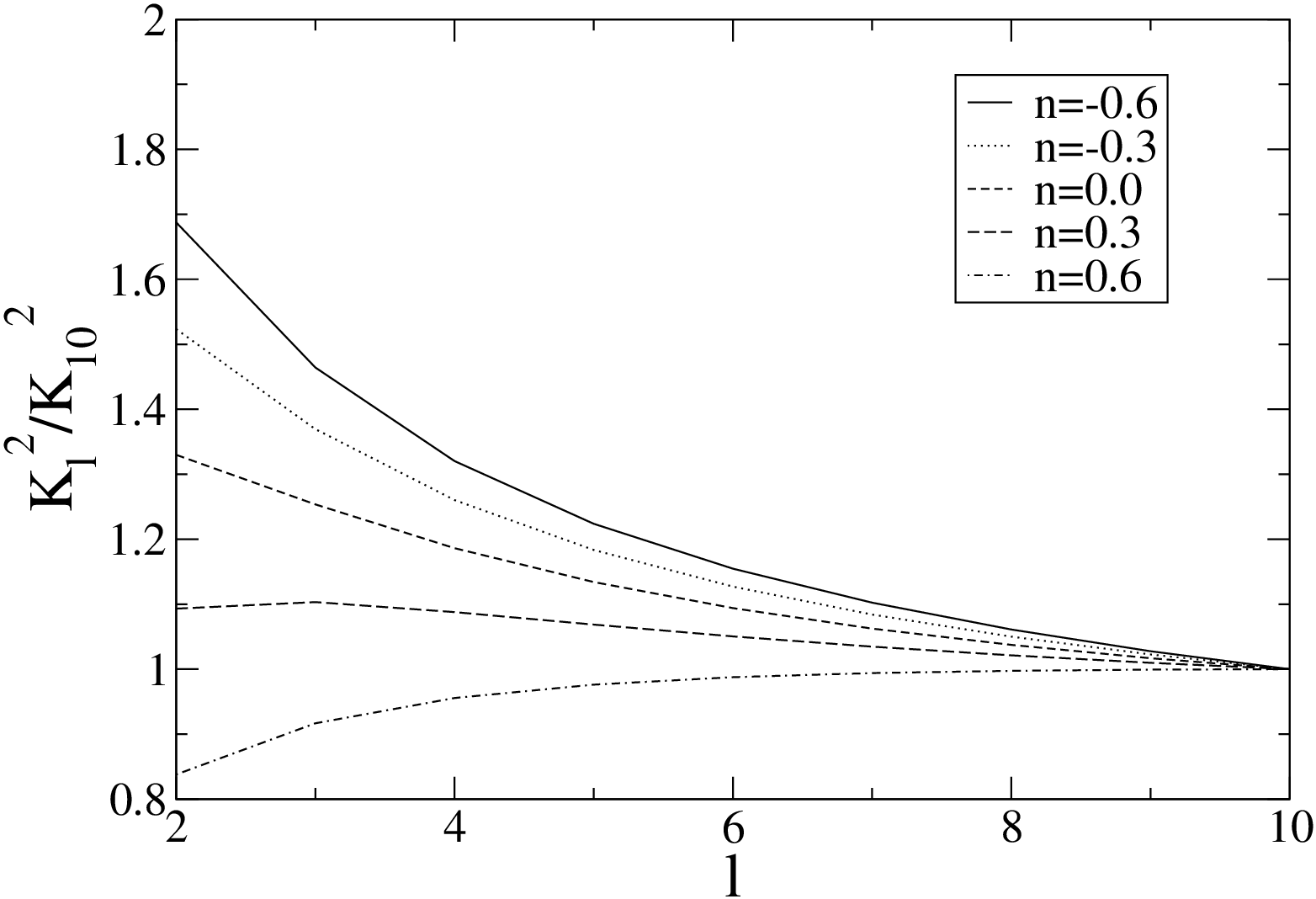}
\includegraphics[width=60mm,angle=-90]{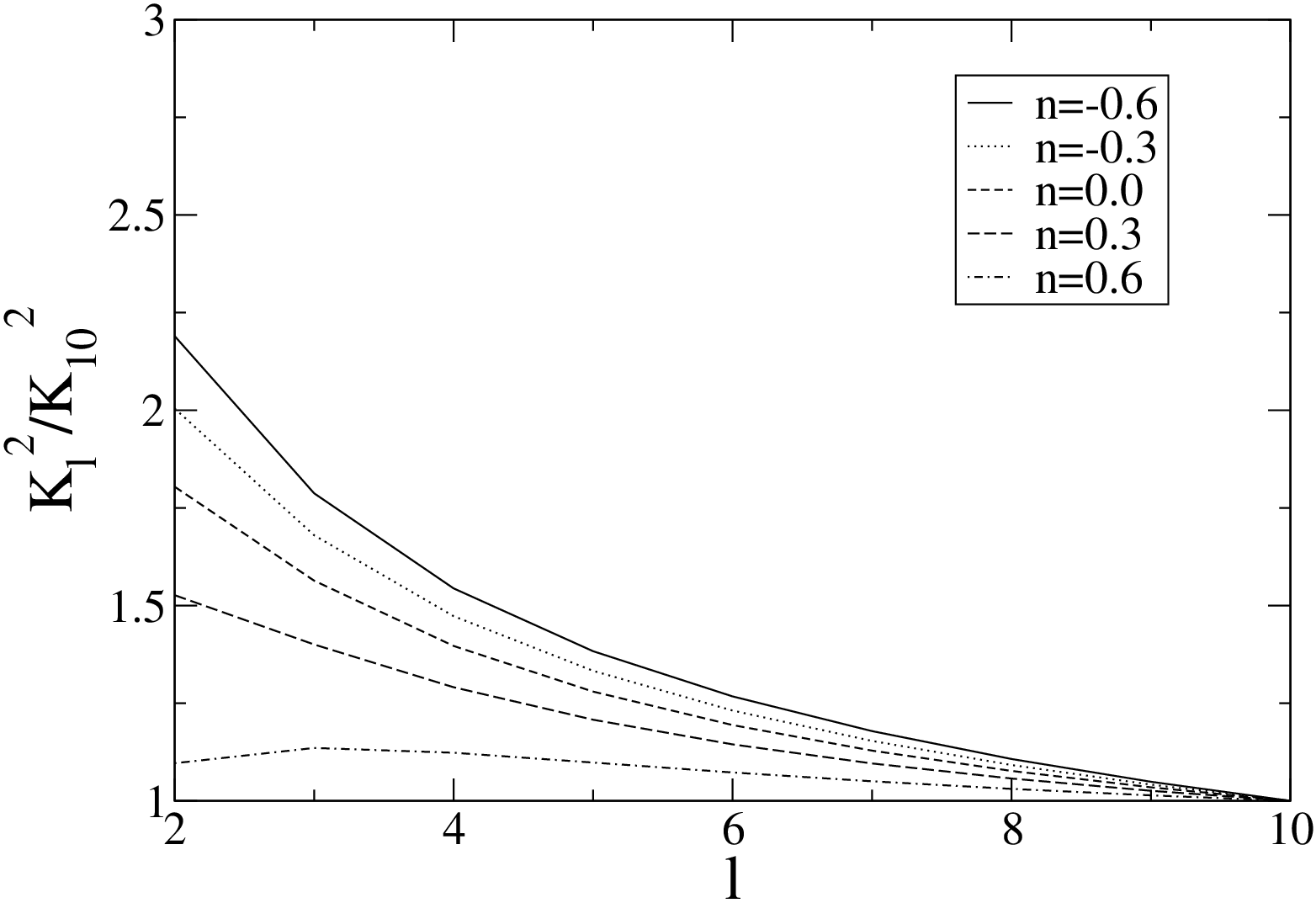}
\caption{$K_l^2/K_{10}^2$ in the MPC model, $q=1$ (left), $q=5$ (right)}
\label{mpcpic}
\end{center}
\end{figure}

We have calculated the ISW effect for different sets of parameters. Even though at large $q$
the linear growth of fluctuations is larger than in the \lcdm model, we find that  the quadrupole
is enhanced for large $q$. In fact, the quadrupole can only be suppressed at small $q$. This is 
demonstrated in Fig. \ref{mpcpic}, where we have plotted the normalized large scale power
for $q=1,\ 5$, $n=-0.6,\ -0.3,\ 0,\ 0.3,\ 0.6$. The figure is plotted for $\Omega_M=0.25$.
From the figure we can see that for small $q$, the quadrupole can be suppressed for certain values of $n$ 
but at large $q$ the quadrupole is more likely to be enhanced.

\section{ISW and linear growth}
Cosmic variance is obviously a significant hindrance when considering the large order multipoles.
Hence, it can be difficult to make observationally significant predictions on the shape of
the power spectrum at large $l$. However, non-standard cosmological evolution does not
only have an effect on the shape but also on the overall normalization of the power
spectrum due to the ISW effect, which can be a useful tool in differentiating between 
different models. As we have seen in the previous section, non-standard evolution
can, in addition to suppressing the low multipoles, also have a significant effect on the
normalization.

From Eq. (\ref{isw}) it is clear that the power spectrum has two contributions: the amplitude
of the primordial fluctuations $A$ and the ISW effect. The CMB observations measure this product
and not the amplitude of the primordial fluctuations directly.  On the other hand, the amplitude 
of the primordial fluctuations is also probed by matter fluctuations. A common
measure of the matter fluctuations is the current value of rms fluctuations on a sphere of 
$8$ Mpc/h, $\sigma_8$. Given the amplitude of primordial fluctuations and a particular 
cosmological model, the value of $\sigma_8$ can be calculated. WMAP \cite{wmap1} results 
indicate that $\sigma_8=0.84\pm 0.04$ (best fit model with a running spectral index). 

Both probes of primordial fluctuation are linear functions of $A$,
$\sigma_8\propto A D,\ C_l\propto A^2 K_l^2$, where $D$ is the current value of the linear growth factor 
in a particular cosmological model. The observational quantitities, $\sigma_8$ and $C_l$, are therefore
related  to the theoretical quantities, $D$ and $K_l^2$ by
\be{iswsigma}
{\sigma_8\over l(l+1) C_l}\propto {D^2\over K_l^2}.
\ee
The evolution of the linear growth factor $D$ in a cosmological model with a general Friedmann
equation is determined by \cite{mgm}
\be{genlinear}
{d^2D\over d\tau^2}+(2+{\dot{\bH}\over \bH^2}){dD\over d\tau}+3 c_1D=0,
\ee
where $\tau=\ln(a)$, $\bH$ is the unperturbed Hubble rate and 
$c_1$ is determined by the expansion
\be{genrhsexp}
3{1+\delta\over\bH^2}\Big((\dot{H}+H^2)-(\dot{\bH}+\bH^2)\Big)
\equiv 3(1+\delta)\sum_{n=1} c_n\delta^n.
\ee
Here $\delta=\rho/\bar{\rho}-1$ is the local density contrast and $H=H(\rho)$ is the perturbed Hubble rate.

\begin{figure}[ht]
\begin{center}
\includegraphics[width=60mm,angle=-90]{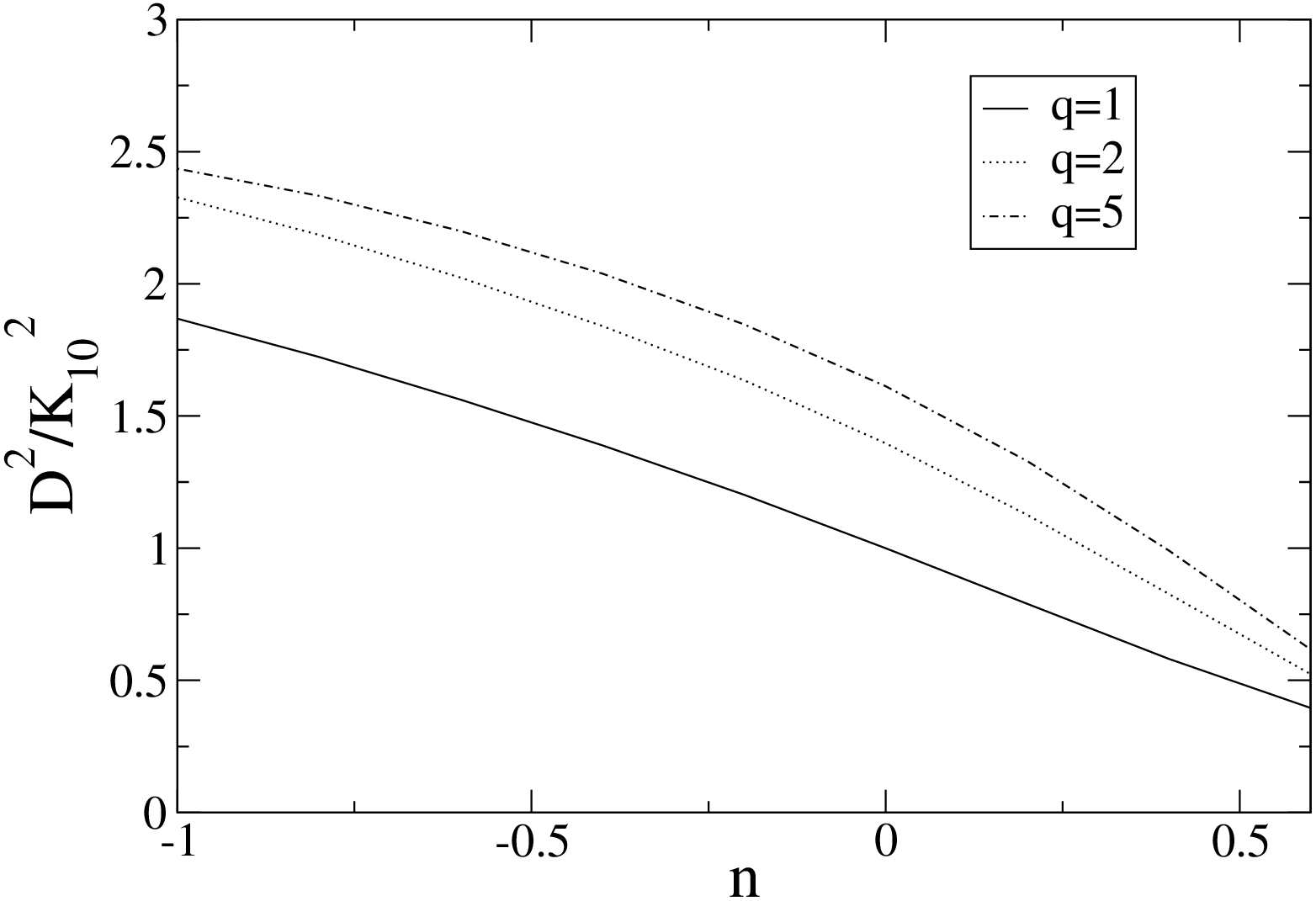}
\includegraphics[width=60mm,angle=-90]{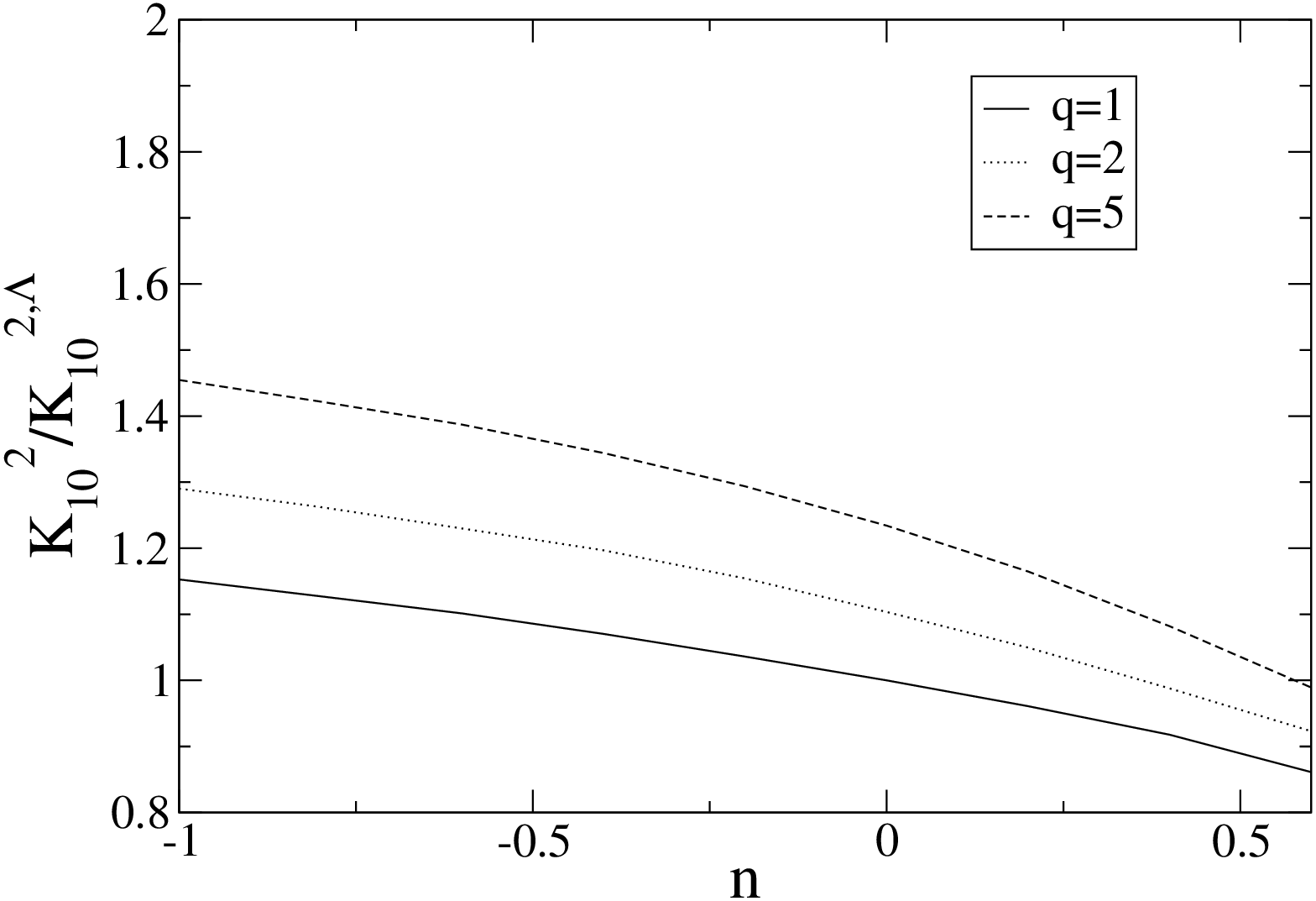}
\caption{$D/K_l^2$ (left) and $K_{10}^2$ (right) relative to \lcdm in the MPC model for
different values of $q$ as a function of $n$}
\label{iswpic}
\end{center}
\end{figure}

Eq. (\ref{genlinear}) is straightforwardly solvable numerically for each cosmological model, along with
the ISW effect. To illustrate, we have plotted $D^2/K_l^2$ at $z=0$ as well as
$K_{10}^2/K_{10}^{2,\Lambda}$ for the MPC model for different values of $q$ and $n$ in Fig. \ref{iswpic}. 
Both figures are normalized to \lcdm values.
From the figures it is evident that the additional information from $\sigma_8$ helps to 
discriminate between models. 

In order to discriminate between dark energy models, it is hence important to
combine CMB observations  with the measurements of $\sigma_8$ from matter surveys, as pointed out in \cite{kunz}.
Measuring the CMB alone cannot tell us what is the actual amplitude of the initial perturbations,
but we must combine it with $\sigma_8$ to sidestep the issue.
Note that since the ISW effect can enhance the initial fluctuations, one can in principle
have less initial power from inflation. The ISW effect can  therefore be also significant
from the point of view of inflationary model building.

As we have seen, the typical situation is where the ISW tends to increase the large-scale 
CMB power leading to a lower normalization and a lower value for $\sigma_8$.  
If the dark energy clusters 
on large scales, there is an extra modification to the matter 
power spectrum for comoving wavenumbers less than 
\begin{equation}
k_Q \sim 10^{-3} \sqrt{(1-w)(2+2w-w\Omega_{\rm m})}\;h\,{\rm Mpc}^{-1},
\label{eq:kq}
\end{equation}
where $w$ is the effective equation of state parameter \cite{ma}.  
The main effect is again to lower $\sigma_8$ compared to standard 
$\Lambda{\rm CDM}$.  This suggests that it should be possible to 
use $\sigma_8$ to discriminate between dark energy models.  
Note, however, that several other effects have a similar impact on 
$\sigma_8$.  For  example, massive neutrinos reduce power on 
comoving wavenumbers greater than 
\begin{equation}
k_{\rm nr} \approx 0.02\left(\frac{m_\nu}{1\;{\rm eV}}\right)^{1/2}
\Omega_{\rm m}^{1/2}\;h\,{\rm Mpc}^{-1},
\label{eq:knr}
\end{equation}
where $m_\nu$ is the common neutrino mass (i.e. we have assumed three 
equal-mass neutrinos).  As long as a neutrino mass as large as $m_\nu \sim 
0.1\;{\rm eV}$ cannot be excluded, it is hard to use the clustering 
amplitude to distinguish between models of dark energy.  
As an illustration, standard $\Lambda{\rm CDM}$ (flat universe with 
$\Omega_{\rm m}=0.3$, scale-invariant adiabatic primordial fluctuations) 
gives $\sigma_8=0.93$ after normalizing to COBE.   With a constant 
equation of state $w=-0.7$ for the dark energy component, and the 
remaining parameters fixed, one gets $\sigma_8 = 0.79$, whereas   
$\Lambda{\rm CDM}$ with a contribution $\Omega_\nu h^2 = 0.005$ from 
massive neutrionos to the dark matter density (corresponding to 
$m_\nu = 0.15\;{\rm eV}$) gives $\sigma_8 = 0.82$.

\section{Discussion and Conclusions}
In this work we have studied the importance of the ISW effect in non-standard cosmologies.
A possible signature is the observed lack of
large scale power in the cosmic microwave background radiation.
Our discussion is relevant to flat universes where the dark energy component does not fluctuate,
like the \lcdm model. Extending the discussion to the case where dark energy also fluctuates
is left for future work.

In the \lcdm model, the ISW effect acts to enhance the fluctuations on large
scales in such a way that effectively, 
$|\Delta^{SW}+\Delta^{ISW}|^2\approx |\Delta^{SW}|^2+|\Delta^{ISW}|^2$.
As we have argued, in the general case this does not have to hold, and the cross term
can be significant. To illustrate the point, we have studied different examples of models where the
ISW effect can reduce the large scale power. For example, we have seen that in an EdS-universe
which undergoes a period of non-standard growth, one can easily suppress the quadrupole
as is shown in Fig. \ref{deltapic}. Unfortunately, in addition to the fact that the EdS-model is not
compatible with the cosmological concordance model, strong suppression also requires that 
the universe evolves in a non-standard way throughout the most of its history since recombination. 
If one further assumes that we should recover the \lcdm behaviour at low redshifts, the situation becomes
worse as one cannot then suppress the quadrupole as much.

Considering a more realistic model, where there is a brief period of non-standard growth
within the \lcdm model, we have seen that one can reduce the large scale power by a rapid
phase where $a\sim 1/t$. This means that the universe contracts briefly.  
The effect on the quadrupole is larger the more recently this period occurs, as Fig. \ref{ddlcdm4} demonstrates.

An obvious omission in this work is that we have not given any particular physical model 
for the unorthodox behaviour of the universe but concentrated simply on whether the 
expansion of the universe can reduce the large scale power. To put this idea into more solid 
footing one should consider models where the universe can go through a rapid 
contraction phase. It is not obvious that one cannot device a model where we can produce 
the observed multipoles exactly, \ie one can consider the inverse problem of going from the 
power spectrum to the evolution of the scale factor. This is most likely an academic question, 
unless one finds that the evolution is not modified too radically from the standard picture. 
Again, we leave these questions for further studies.

In addition to possibly alleviating the problems associated with the lack of large scale power,
one can potentially also use the ISW to differentiate between different dark energy models
in a way that is independent of the  amplitude of primordial fluctuations.
Combined with CMB independent observations of $\sigma_8$, the ISW effect gives
constraints on dark energy models. Interestingly, since we only observe the CMB filtered
through the ISW effect, one can speculate whether the amplitude primordial fluctuations
can be in fact lower than what is typically assumed.

The integrated Sachs-Wolfe effect is a useful tool for cosmology. It probes the whole history
of cosmological evolution from recombination until the present time.
It can act to enhance, but also reduce, power on large scales. 
Both effects can be important for cosmology.

\begin{acknowledgments}
TM would like to thank E. Gazta\mytilde{n}aga for fruitful discussions and
Institut d'Estudis Espacials de Catalunya (IEEC) for hospitality.
\end{acknowledgments}


\end{document}